\documentclass[twocolumn,superscriptaddress,pra,aps]{revtex4}
\usepackage{mathptmx}
\usepackage{subfigure}
\usepackage{psfrag,graphicx}
\usepackage{dcolumn}
\usepackage{amsmath,amssymb}
\usepackage{bm}
\usepackage{color}
\usepackage{latexsym}
\usepackage{epstopdf}
\usepackage{color}
\usepackage[english]{babel}
\usepackage{latexsym}
\usepackage{psfrag,graphicx}
\usepackage{subfigure}
\usepackage{amsmath}
\usepackage{amssymb}
\usepackage{amsfonts}
\usepackage{bm}
\usepackage{natbib}
\usepackage{epstopdf}
\usepackage{color}
\usepackage{appendix}
\usepackage[colorlinks=true,citecolor=myblue,linkcolor=myred]{hyperref}

\definecolor{mygrey}{gray}{0.35}
\definecolor{myblue}{rgb}{0.2,0.2,0.8}
\definecolor{myzard}{cmyk}{0,0,0.05,0}
\definecolor{mywhite}{rgb}{1,1,1}
\definecolor{mywhite}{rgb}{1,1,1}
\definecolor{myred}{rgb}{1,0.,0.3}
\def\be{\begin{equation}}
\def\ee{\end{equation}}
\def\bse{\begin{subequations}}
\def\ese{\end{subequations}}
\def\ba{\begin{align}}
\def\enda{\end{align}}
\def\bi{\begin{itemize}}
\def\ei{\end{itemize}}

\newcommand{\ket}[1]{|#1\rangle}

\def\ket#1{\vert #1 \rangle}

\def\to{\rightarrow}
\def\fromto{\leftrightarrow}

\def\sech{\textrm{\,sech}}
\def\ket#1{\vert #1\rangle}
\def\H{\mathbf{H}}
\def\W{\mathbf{W}}
\def\eps{\lambda}
\def\vec{\chi}
\def\sech{\text{sech}\,}
\def\sec#1{\textbf{#1.}}

\begin{document}

\title{Optically-Induced Highly-Efficient Detection and Separation of Chiral Molecules through Shortcuts to Adiabaticity}

\author{Nikolay V. Vitanov}
\affiliation{Department of Physics, St Kliment Ohridski University of Sofia, James Bourchier 5 blvd, BG-1164 Sofia, Bulgaria}
\author{Michael Drewsen}
\affiliation{Department of Physics and Astronomy, Aarhus University, DK-8000 Aarhus C, Denmark}

\date{\today }

\begin{abstract}
A highly-efficient method for optical detection and separation of left- and right-handed chiral molecules is presented.
The method utilizes a closed-loop three-state system in which the population dynamics depends on the phases of the three couplings.
Due to the different signs of the coupling between two of the states for the opposite chiralities and the phase sensitivity of the scheme, the population dynamics is chirality-dependent.
By using concepts from the ``shortcuts to adiabaticity'' approach, one can achieve 100\% contrast in state population transfer between the two enantiomers, which can be probed by light-induced fluorescence in the case of large ensembles or through resonantly-enhanced multiphoton ionization at the single molecular level.
\end{abstract}

\maketitle

\sec{Introduction}
Ever since its discovery by Pasteur in 1848 \cite{Pasteur1848} chirality has played an important role in chemistry, biotechnologies and pharmaceutics due to its significance to living matter.
Moreover, precise control of the rotational populations of chiral molecules opens a plethora of new possibilities, e.g. studies of chemical processes and parity violation in chiral molecules \cite{Quack2002,Quack2008}.
Yet, the detection and separation of racemic mixtures into their enantiomers, i.e. molecules with opposite left (L) and right (R) handedness, referred to as chiral resolution,  is one of the outstanding challenges in chemistry \cite{Knowles2002}.
Conventionally, it is achieved by sophisticated, time-consuming and expensive chemical techniques, e.g. crystallization, derivatization, kinetic resolution, and chiral chromatography \cite{Ahuja2011}.

\begin{figure}[tb]
\begin{tabular}{cc}
\includegraphics[width=0.40\columnwidth]{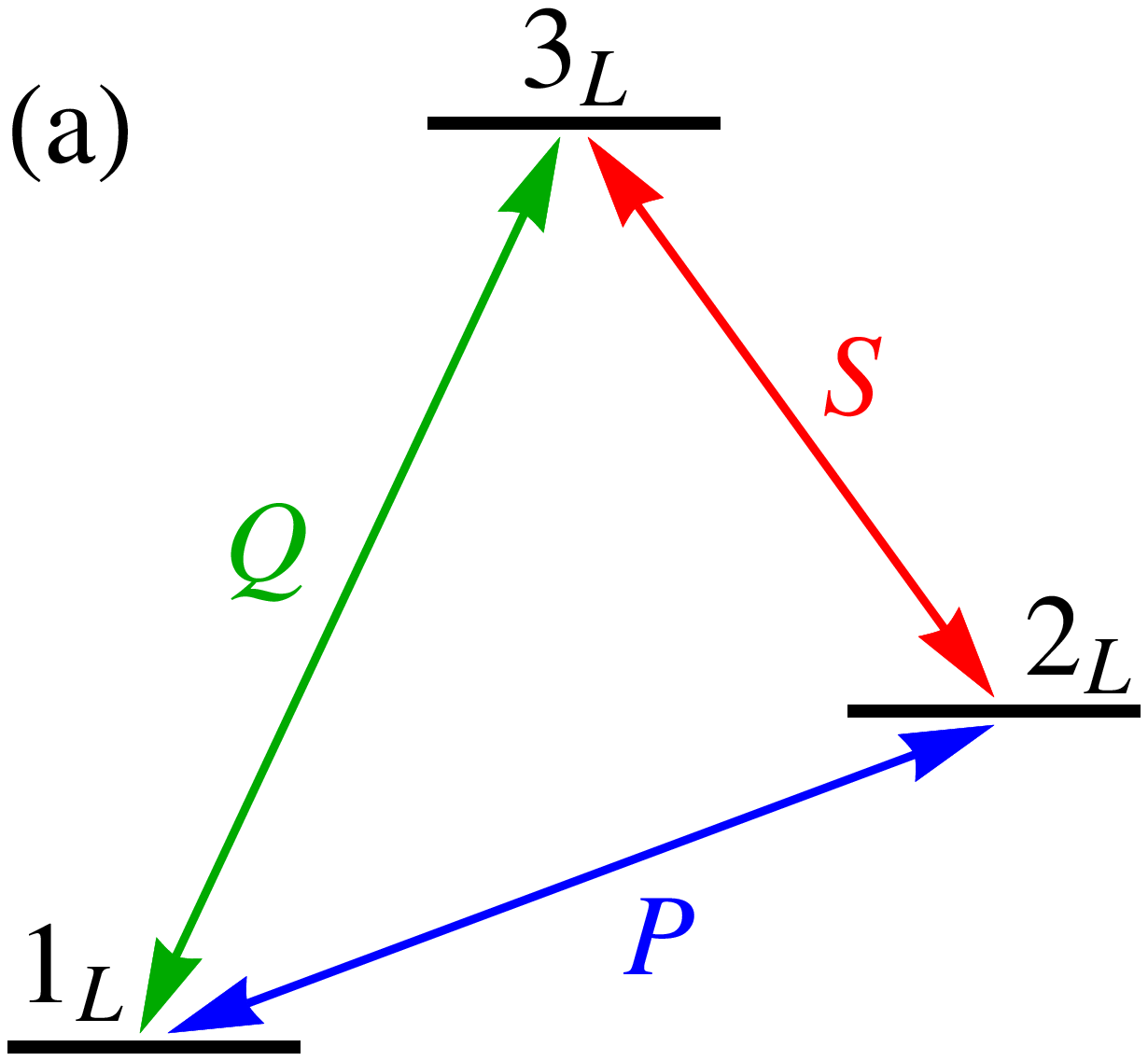} & \hspace{10mm}
\includegraphics[width=0.40\columnwidth]{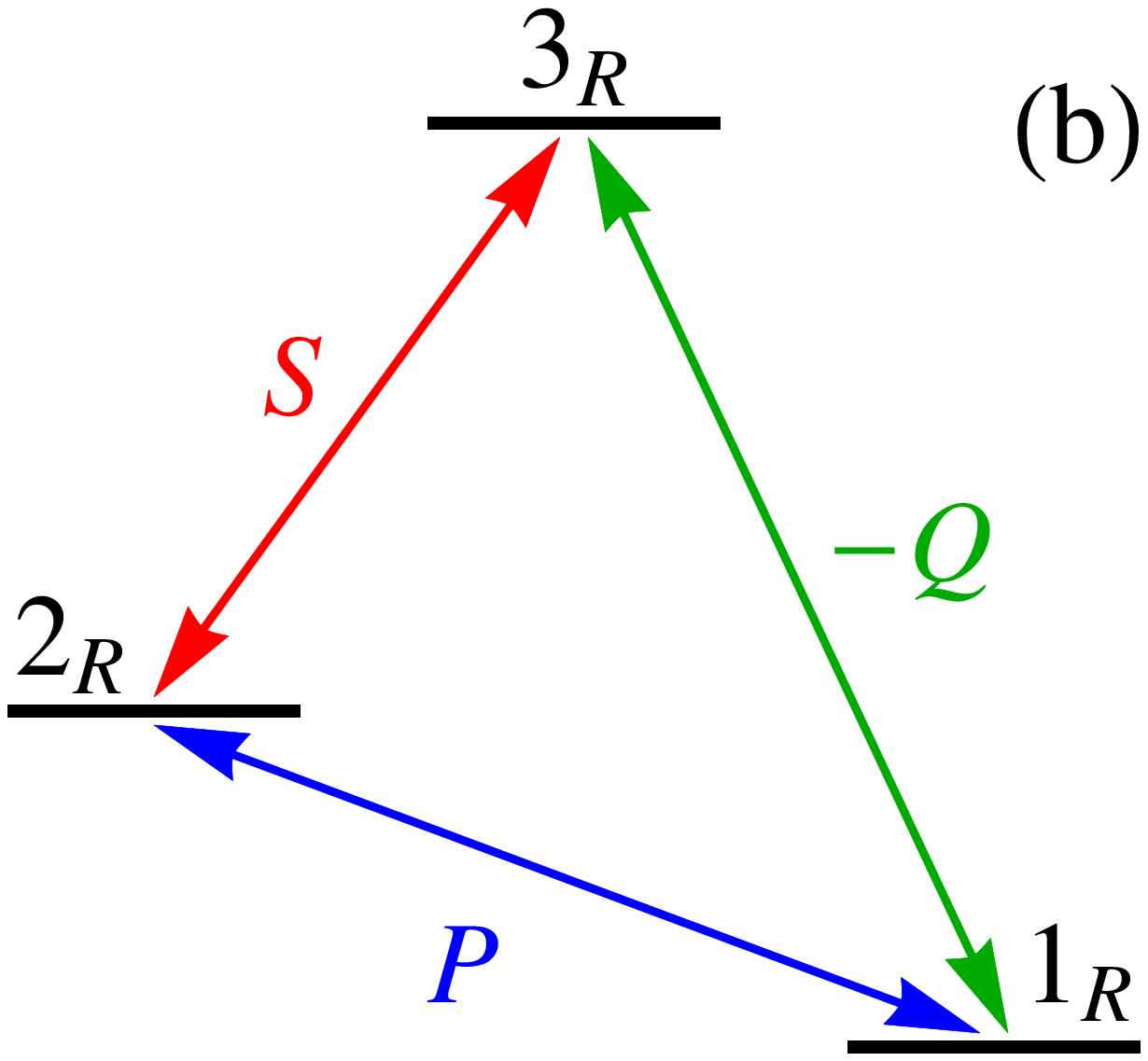}
\end{tabular}
\caption{Comparison of the coupling schemes between three discrete energy states in molecules with L (a) and R (b) handedness.
The only difference is in the sign of the coupling $Q$ on the transition $\ket{1} \fromto \ket{3}$.
}
\label{fig:chiral-scheme}
\end{figure}

As an alternative, optical (termed ``chiroptical'') methods have been developed and they offer some promising features.
They use the fact that the mirror symmetry of enantiomers can be broken using circularly polarized electromagnetic fields \cite{Berova2012}.
Well established chiroptical methods include optical rotary dispersion \cite{Berova2012}, circular dichroism \cite{Berova2000}, vibrational circular dichroism \cite{Nafie1976,Nafie2011}, and Raman optical activity \cite{Baron2004,Nafie2011}.
Circularly polarized light interacts with chiral molecules via the (weak) magnetic-dipole interaction \cite{Salam1997,Salam1998}.
Alternative methods for achieving enantio-selectivity have been proposed with linearly polarized light that use the far stronger electric-dipole interaction with the field \cite{Fujimura1999,Fujimura2000,Hoki2001,Hoki2002,Gonzales2001,Kroner2003}.
What makes chiral separation feasible in such schemes is the sign difference in some of the transition dipole moments of the L and R enantiomers (see Fig.~\ref{fig:chiral-scheme}),
 which can be mapped onto population difference by subjecting the chiral molecules to an appropriately timed and phased set of external electromagnetic fields.

In a complementary approach, Shapiro and co-workers \cite{Shapiro2000,Brumer2001,Gerbasi2001,Frishman2003,Kral2001,Kral2003,Thanopulos2003,Gerbasi2004} proposed to detect and separate enantiomers by using concepts from the adiabatic passage techniques \cite{Vitanov2017}, which offer robustness to various experimental errors.
They considered a four-state closed-loop double-$\Lambda$ or diamond configuration \cite{Shapiro2000,Brumer2001,Gerbasi2001,Frishman2003},
and a closed-loop three-state system driven by three delayed but overlapped laser fields (P, S, and Q) \cite{Kral2001,Kral2003,Thanopulos2003,Gerbasi2004}, see Fig.~\ref{fig:chiral-scheme}.
In addition, nonzero detunings are used, which generate avoided crossings between the adiabatic states.
For opposite handedness, the avoided crossings occur between different adiabatic states, thereby driving the population evolution along different paths and directing the population toward different final states.

While chiral signals have not yet been demonstrated through the optical techniques based on the electric-dipole moment, chiral resolution in the gas phase by electromagnetic fields has been demonstrated using chirality-sensitive microwave spectroscopy \cite{Domingos2018}.
Here, three-wave mixing in a cooled gas cell or in a supersonic molecular beam \cite{Patterson2013n,Patterson2013prl,Shubert2014,Shubert2016} was combined with the newly developed broadband chirped-pulse rotational spectroscopy \cite{Brown2008,Park2016}.
Chiral separation can furthermore be achieved by applying three resonant pulsed fields with appropriate phases on all three transitions which lead to different enantio-selective population distributions \cite{Li2008}, as has very recently been demonstrated experimentally in both buffer gas cell \cite{Eibenberger2017} and supersonic expansion \cite{Perez2017}.

\begin{figure}[tb]
\includegraphics[width=0.80\columnwidth]{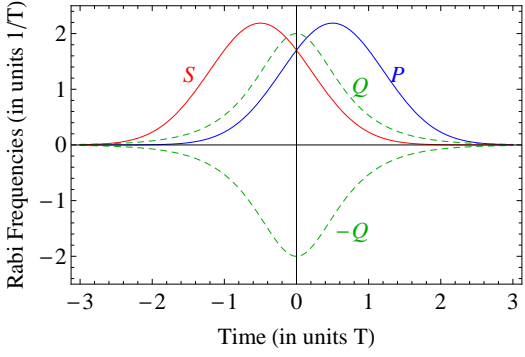}
\caption{
Timing of the P, S and Q couplings for Gaussian pulse shapes of the P and S fields.
}
\label{fig:Gaussian}
\end{figure}

In this Letter, we follow the original idea of Shapiro and co-workers and propose a different scheme, which is simpler and much faster, and its efficiency is essentially 100\%.
It still requires delayed pulses with some mutual overlap (see Fig.~\ref{fig:Gaussian}), but it works on resonance, i.e. at null detunings.
Very importantly, the required pulse areas are only of the order of $\pi$, which means that the scheme is very fast.

Our scheme combines the loop linkage of Shapiro and co-workers with an idea reminiscent to what is nowadays known as ``shortcuts to adiabaticity''.
It was first proposed by Bergmann and co-workers \cite{Unanyan1997}, who suggested to use the Q pulse to cancel the nonadiabatic coupling in the well known and widely used technique of stimulated Raman adiabatic passage (STIRAP) and ensure perfect adiabaticity with moderate pulse areas.
Essentially the same idea was proposed later by Demirplak and Rice \cite{Demirplak2003} who used the term ``counterdiabatic field'',
 and by Chen \emph{et al.} \cite{Chen2010} who nailed the very appealing term ``shortcut to adiabaticity''.

\sec{Mechanism of enantiomer detection and separation}
Because of its interferometric nature, the closed-loop scheme is phase-sensitive.
It is exactly this phase sensitivity which is used here to separate the two enantiomers because the overall phase in the loop linkage differs for the two enantiomers by $\pi$.
The key idea of our scheme is that for one enantiomer, e.g. L-handed, the Q field works as a ``shortcut to adiabaticity'' for it cancels the nonadiabatic coupling and induces perfect population transfer between states $\ket{1}$ and $\ket{3}$, with probability $P_{1\to 3} = 1$.
For the opposite R handedness, the same Q field acts oppositely due to the different sign of the coupling, and it doubles rather than cancels, the nonadiabatic coupling.
For specific values of the pulse areas, the population transfer $\ket{1}\to\ket{3}$ is cancelled completely: $P_{1\to 3} = 0$.
Therefore, L or R handedness is differentiated by measuring the population in state $\ket{3}$ alone: 0 for one handedness and 1 for the opposite handedness.

We consider a three-state system in which all three transitions are driven simultaneously, thereby forming a phase-sensitive interferometric linkage pattern, as shown in Fig.~\ref{fig:chiral-scheme}.
The Hamiltonian of this system reads
\be\label{H}
\H^{L,R} = \frac12 \left[\begin{array}{ccc}
 0 & \Omega_p & \pm\Omega_q e^{i\phi} \\
 \Omega_p^\ast & 0 & \Omega_s \\
 \pm\Omega_q e^{-i\phi} & \Omega_s^\ast & 0 \\
\end{array}\right] ,
\ee
where the superscripts $L$ and $R$ denote the handedness.
$\Omega_x(t)$ with $x=p,s,q$ is the Rabi frequency quantifying the coupling for the P, S or Q transition.
The only difference between L and R handedness is in the sign of the Q coupling, the $+$ ($-$) sign being for L (R) handedness.
The Q coupling has a phase $\phi$, which, following Ref.~\cite{Unanyan1997} will be set to $\pi/2$.
Without loss of generality we assume hereafter that $\Omega_p$, $\Omega_s$, and $\Omega_q$ are real and positive.

First, let us assume that the Q coupling is zero, i.e.
\be\label{H0}
\H_0 = \frac12 \left[\begin{array}{ccc}
 0 & \Omega_p & 0 \\
 \Omega_p & 0 & \Omega_s \\
 0 & \Omega_s & 0
\end{array}\right] .
\ee
The three eigenvalues of this Hamiltonian are
$\eps_- = -\Omega$,  $\eps_0 = 0$, $\eps_+ = \Omega$,
where $\Omega = \sqrt{\Omega_p^2 + \Omega_s^2}$.
The corresponding eigenvectors of $\H_0$ are
\bse\label{eigenvectors}
\begin{align}
\ket{\vec_-}  &= \frac{\sin \theta \ket{1} - \ket{2} + \cos \theta \ket{3}}{\sqrt{2}}  , \\
\ket{\vec_0} &= \cos \theta \ket{1} - \sin\theta \ket{3} , \\
\ket{\vec_+}  &= \frac{\sin \theta \ket{1} + \ket{2} + \cos \theta \ket{3}}{\sqrt{2}} ,
\end{align}
\ese
with
$\tan\theta(t) = \Omega_p(t) / \Omega_s(t)$.
In STIRAP, the pump and Stokes pulses are delayed to each other, with the Stokes pulse coming first \cite{Vitanov2017}.
Therefore, $\theta(t)$ changes from 0 initially to $\pi/2$ in the end and hence the eigenvector $\ket{\vec_0}$ behaves as $\ket{1}$ initially and $-\ket{3}$ at the end, thereby providing an adiabatic link between these two states.
If the system starts in state $\ket{1}$ initially and evolves adiabatically, then it will remain in the eigenstate $\ket{\vec_0}$ at all times and will end up in state $\ket{3}$.

We use the eigenstates \eqref{eigenvectors} to form the transformation matrix
\be\label{W}
\W = \left[\begin{array}{ccc}
 \frac{\sin \theta }{\sqrt{2}} & \cos \theta & \frac{\sin \theta }{\sqrt{2}} \\
 -\frac{1}{\sqrt{2}} & 0 & \frac{1}{\sqrt{2}} \\
 \frac{\cos \theta }{\sqrt{2}} & -\sin \theta  & \frac{\cos \theta }{\sqrt{2}}
\end{array}\right],
\ee
and we switch the Q coupling on.
The full Hamiltonian \eqref{H} (with $\phi=\pi/2$) is transformed in the basis of the eigenstates \eqref{eigenvectors} of $\H_0$ as $\H_a = \W^T \H \W - i \W^T \dot \W $.
Explicitly,
\be\label{Ha}
\H_a^{L,R} =  \left[\begin{array}{ccc}
 -\frac12\Omega  & \frac{i}{\sqrt{2}} (\dot\theta \mp \frac12\Omega_q) & 0 \\
 - \frac{i}{\sqrt{2}} (\dot\theta \mp \frac12\Omega_q) & 0 & - \frac{i}{\sqrt{2}} (\dot\theta \mp \frac12\Omega_q) \\
 0 &  \frac{i}{\sqrt{2}} (\dot\theta \mp \frac12\Omega_q) & \frac12\Omega
\end{array}\right] ,
\ee
with the upper (lower) sign being for L (R) handedness.
Here the choice of the phase $\pi/2$ of the Q-coupling in Eq.~\eqref{H} becomes clear: the Q-coupling adds to or subtracts from the nonadiabatic coupling $\dot\theta$.
We have achieved something very important: the sign difference in the L and R couplings for the Q transition is mapped onto different magnitudes of the couplings between states $\ket{\vec_0}\fromto \ket{\vec_-}$ and $\ket{\vec_0}\fromto \ket{\vec_+}$.
It is this difference that is used in our method to separate the two enantiomers.

If we choose (see Fig.~\ref{fig:Gaussian})
\be\label{Q}
\Omega_q(t) = 2\dot\theta (t),
\ee
then the off-diagonal elements in $\H_a^L$ will vanish and the system will be locked in its initial state $\ket{\vec_0}$ as no transition in this basis can occur.
On the contrary, the off-diagonal elements in $\H_a^R$ will double in magnitude and will induce transitions $\ket{\vec_0}\to \ket{\vec_-}$ and $\ket{\vec_0}\to \ket{\vec_+}$ with some probability
\footnote{We note that if we choose, instead, $\Omega_q(t) = -2\dot\theta (t)$, then the opposite will happen: the off-diagonal elements of $\H_a^L$ will double in magnitude, while those of $\H_a^R$ will vanish.}.
Because for the P and S pulse order in Fig.~\ref{fig:Gaussian} the dark state  $\ket{\vec_0}$ is equal to state $\ket{1}$ in the beginning and to state $-\ket{3}$ in the end, then remaining in the dark state $\ket{\vec_0}$ for the L handedness implies complete population transfer from state $\ket{1}$ to state $\ket{3}$.
On the contrary, if the dark state $\ket{\vec_0}$ is depleted due to the enhanced nonadiabatic coupling for the R handedness, then no population will be found in state $\ket{3}$ in the end.
Therefore, if the three-state system is initially in state $\ket{1}$ then the different signs of the Q-coupling for L and R handedness will map onto different population distributions at the end of the interaction: monitoring the population of state $\ket{3}$ alone will tell us if the molecule is L- or R-handed.

It is important to note that, contrary to STIRAP \cite{Vitanov2017}, here the aim is \emph{not} adiabatic evolution because in the adiabatic limit the nonadiabatic coupling ($\propto \dot\theta$) will be suppressed by the large pulse areas and complete population transfer will take place for both L and R handedness, thereby rendering the difference in the coupling magnitudes irrelevant.
Here it is crucial to have a significant nonadiabatic coupling $\dot\theta (t)$, so that the two enantiomers behave very differently.
It turns out that this condition is fulfilled for moderately small pulse areas (of the order of $\pi$), which implies that the proposed method is much faster than the earlier adiabatic scenarios.

\sec{Maximum contrast in population transfer}
From an experimental point of view, ideally, for one handedness (L in this example) we want to have transition probability 1 to state $\ket{3}$, and for the opposite handedness (R in this example) transition probability 0.
This would create a maximum contrast in the signals for the two enantiomers.
The value 1 for the L handedness is guaranteed by the choice of Eq.~\eqref{Q}.
The value 0 for the R handedness can occur for a particular choice of the experimental parameters, as explained below.

\begin{figure}[tb]
\includegraphics[width=0.80\columnwidth]{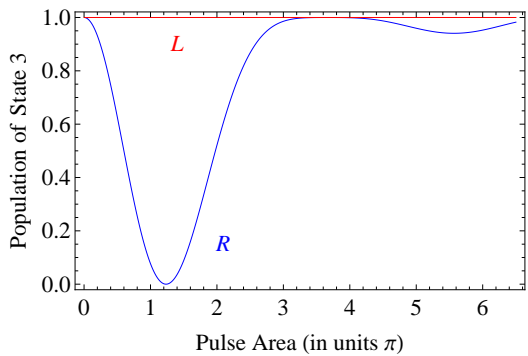}
\caption{
Population $P_3$ of state $\ket{3}$ vs the pulse area $A$ of the P and S couplings for Gaussian pulses with delay $\tau=T$, and the Q-coupling is given by Eq.~\eqref{Q}.
The population $P_3$ for L handedness is exactly equal to 1 due to the ``shortcut to adiabaticity'', whereas it varies for R handedness due to the nonzero nonadiabatic coupling and approaches 1 as the pulse area $A$ increases because the adiabaticity improves.
The largest contrast between the signals occurs for $A\approx 1.234\pi$, where the R signal vanishes.
}
\label{fig:Gaussian-P3}
\end{figure}

For the commonly used Gaussian pulse shapes, $\Omega_p(t) = \Omega_0 e^{-(t-\tau/2)^2/T^2}$ and $\Omega_s(t) = \Omega_0 e^{-(t+\tau/2)^2/T^2}$, the Q-pulse is $\Omega_q(t) = \pm 2(\tau/T) \sech(2\tau t/T^2)$.
These three pulse shapes are shown in Fig.~\ref{fig:Gaussian}.
For a fixed pulse delay $\tau$, complete population depletion of the adiabatic state $\ket{\vec_0}$ takes place for a special value of the peak P and S Rabi frequency $\Omega_0$, or equivalently, for a special value of the P and S pulse area $A = \Omega_0 T \sqrt{\pi}$.
For example, for $\tau/T = 0.6$, 0.8, 1.0, and 1.2, these values are (derived numerically) $A \approx 0.891\pi$, $1.035\pi$, $1.234\pi$, and $1.510\pi$.
For these pairs of pulse delays and pulse areas, and given the condition \eqref{Q} for the Q-coupling, the application of the P, Q and S pulses of Fig.~\ref{fig:Gaussian} will result in complete population transfer $\ket{1} \to \ket{3}$ for L handedness ($P_3=1$), whereas zero population will reach state $\ket{3}$ for the opposite R handedness ($P_3=0$).
Figure \ref{fig:Gaussian-P3} illustrates these findings.

It is important to note that as long as the Q-coupling has a relative phase of $\pm\pi/2$ there will be a difference between the magnitudes of the couplings for the L and R handedness, even if the condition \eqref{Q} is not precisely fulfilled, and the population distribution (and the ensuing signals) will differ.
However, condition \eqref{Q} is still desirable for it delivers the highest contrast between the L and R signals.
Still, because at the point where $P_3=0$ this population has a minimum, there is some interval of pulse area values wherein the population $P_3$ is very small and the contrast is very high, see Fig.~\ref{fig:Gaussian-P3}.

\begin{figure}[tb]
\includegraphics[width=0.80\columnwidth]{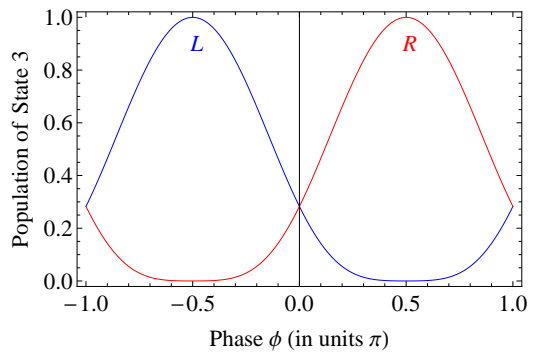}
\caption{
Population $P_3$ of state $\ket{3}$ vs the phase $\phi$ of the Q field for L and R handedness.
The P and S pulses are Gaussian, with a delay $\tau=T$ and pulse areas $A = 1.234\pi$.
The Q-coupling is given by Eq.~\eqref{Q}.
The largest contrast is obtained for $\phi=\pi/2$ and $\phi=3\pi/2$ when one of the population is 1 and the other is 0.
For $\phi=0$ and $\pm\pi$, the two enantiomers produce the same signal and the contrast vanishes.
}
\label{fig:chirality-phase}
\end{figure}

Figure \ref{fig:chirality-phase} shows the expected variations of the L and R signals from state $\ket{3}$ versus the phase $\phi$ of the Q field, which has been assumed to be $\pi/2$ so far.
For $\phi=\pi/2$, the L signal is 1 and the R signal is 0, as expected.
Conversely, for $\phi=-\pi/2$, the L signal is 0 and the R signal is 1.
As in Fig.~\ref{fig:Gaussian-P3}, there are some ranges of phases around the perfect values $\pm\pi/2$ wherein the contrast remains very high.
For phase $\phi = 0, \pm\pi$ the L and R signals are equal and the contrast vanishes completely: the enantiomers are indistinguishable.

So far we assumed Gaussian pulse shapes for the P and S pulses.
Our method is not restricted to these pulse shapes.
Another example is presented in the Supplementary Material for an exactly soluble analytic model, which allows us to derive the transition probability explicitly.

\sec{Enantiomer detection and separation}
For larger ensembles of molecules, the efficiency and contrast in the enantiomer-specific state transfer process could be measured through light-induced fluorescence (LIF) from molecules in state $\ket{3}$, while separation of the enantiomers could be achieved by applying a resonantly-enhanced multiphoton ionization (REMPI) scheme to selectively ionize the enantiomer ending up in state $\ket{3}$, and by an electric field remove the ionized molecules.
Together with the high efficiency in detecting single ions either by an electron-multiplier detector or an ion trap \cite{Hojbjerre2008}, the REMPI method furthermore enables the ultimate detection of a single specific enantiomer molecule.
In case the original chirality of the neutral molecule is preserved after REMPI, the capture of this ion in an ion trap already containing a single laser cooled atomic ion, one can, actually, produce a single chiral molecular ion which is both trapped and cooled to low translational temperatures.
In some cases, one should even be able to produce the chiral molecule in a specific internal state by REMPI, as has been demonstrated with with $N_2^+$ molecules \cite{Tong2010}.
This could enable novel studies of chemical processes and parity violation in chiral molecules \cite{Quack2002,Quack2008}.

The potential advantage of the presented laser based scheme as compared to microwave schemes lie partly in its ability to address states more widely separate in energy, which makes state dependent ionization easier, partly in prospect of spatial very localized interactions through applying crossed laser beams, which is an asset when the goal is to create, trap and conduct experiment with a single chiral molecular ion produced from an initial neutral molecule sample.

\sec{Conclusions}
This paper presented a method for efficient optical detection and separation of chiral molecules.
The main step of preparing the molecules for chiral selection and detection is a closed-loop three-state system driven by three resonant external fields.
All three couplings are the same in magnitude but one of them (Q) has a different sign for L- and R-handed molecules.
By using a suitable pulse delay, as in the STIRAP process, suitable pulse areas, and a phase shift of $\pi/2$ of the Q field, and using concepts from the ``shortcuts to adiabaticity'' method, one can map the sign difference in the Q couplings to different transition probabilities, and hence populations, for the L- and R-handed molecules.

Because our method populates different states in the end of the process, in the case of larger ensembles (e.g., molecules in a cold supersonic beam) it allows the spatial separation of the L- and R-handed molecules in an ensemble, e.g. by REMPI in the presence of an electric field.
The REMPI technique furthermore enables the detection of a single molecule of a particular enantiomer, and potentially constitutes the staring point for the production and investigation of a single molecular ion with a specific chirality.

NVV acknowledges support from the European Commission Horizon-2020 Flagship on Quantum Technologies project 820314 (MicroQC).
MD acknowledges support from the European Commission  FET Open TEQ,  the Villum Foundation and the Sapere Aude Initiative from the Independent Research Fund Denmark.


\end{document}


\title{Supplementary information for the article ``Optically-Induced Highly-Efficient Detection and Separation of Chiral Molecules through Shortcuts to Adiabaticity''}
\author{by Nikolay V. Vitanov and Michael Drewsen}
\maketitle

In addition to the Gaussian pulse shapes considered in the main text, we present here an exactly soluble analytic model.
This analytic solution presents an explicit expression for the transition probability, which allows one to easily analyze its properties.
This model demonstrates that the proposed method for chirality resolution is not limited to the Gaussian pulse shapes but is generally applicable.

\begin{figure}[tb]
\includegraphics[width=0.85\columnwidth]{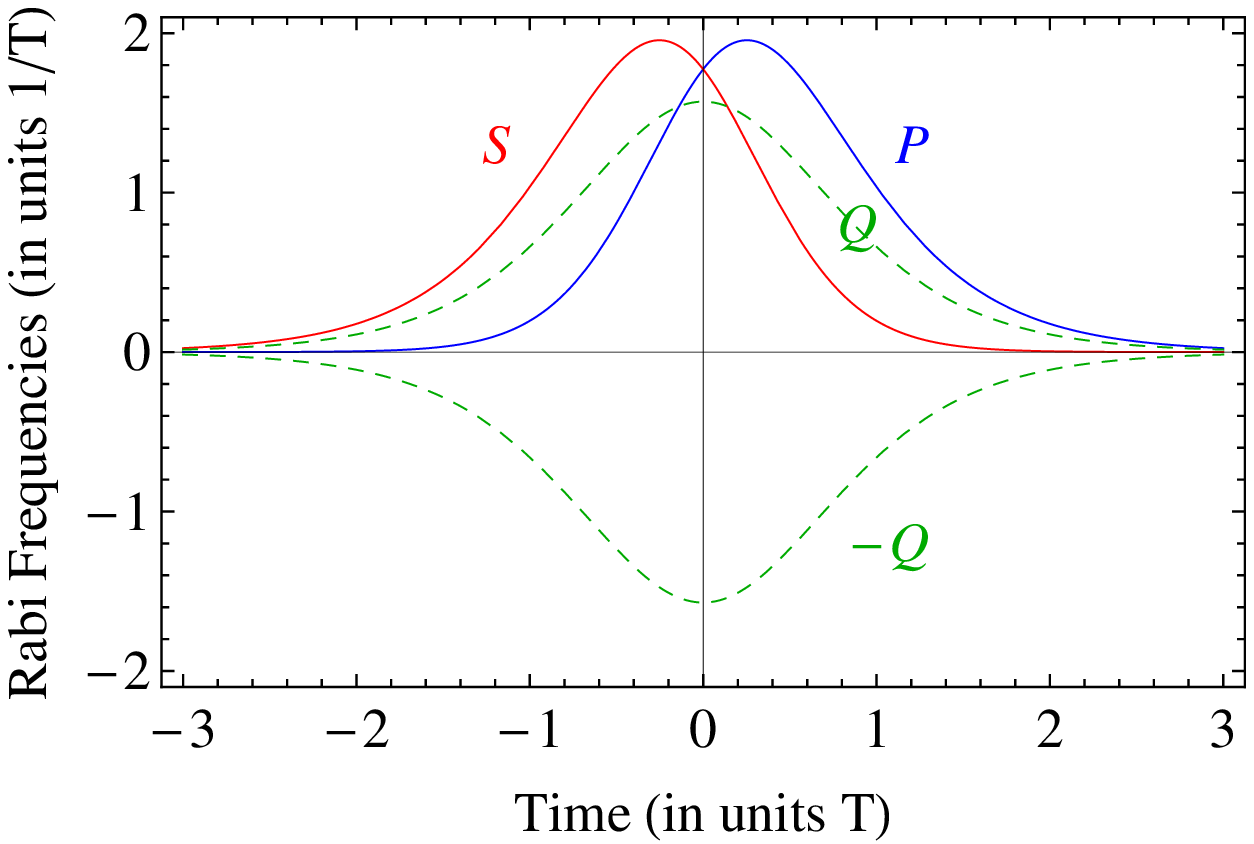}
\caption{
Pulse shapes of the P, S and Q couplings for the analytic model \eqref{shapes-sech}.
}
\label{fig:shapes-sech}
\end{figure}

The pulse shapes in the analytic model are \cite{Vitanov1997}
\bse\label{shapes-sech}
\begin{align}
\Omega_p(t) &= \Omega(t) \sin\theta(t), \\
\Omega_s(t) &= \Omega(t) \cos\theta(t), \\
\Omega_q(t) &= 2 \dot\theta(t).
\end{align}
\ese
with
$\Omega(t) = \Omega_0 \sech^2 (t/T)$ and $\theta(t) = \arctan[\Omega_p(t) / \Omega_s(t)] = [\tanh (t/T) + 1] \pi/4$.
Then $\dot\theta(t) = (\pi/4T) \sech^2(t/T)$ and $\Omega_q(t) = 2 \dot\theta(t) = (\pi/2T) \sech^2(t/T)$.
These pulse shapes are plotted in Fig.~\ref{fig:shapes-sech}.
The pulse area of each of the P and S pulses is $A = (4/\pi) \Omega_0 T$ and the pulse area of the Q-field is $\pi$.
This model has been solved in Ref.~\cite{Vitanov1997} in the absence of the Q-field, $\Omega_q(t) = 0$.
Here we extend the solution to nonzero $\Omega_q(t)$.

The Hamiltonian in the adiabatic basis
\be\label{Ha}
\H_a^{L,R} =  \left[\begin{array}{ccc}
 -\frac12\Omega  & \frac{i}{\sqrt{2}} (\dot\theta \mp \frac12\Omega_q) & 0 \\
 - \frac{i}{\sqrt{2}} (\dot\theta \mp \frac12\Omega_q) & 0 & - \frac{i}{\sqrt{2}} (\dot\theta \mp \frac12\Omega_q) \\
 0 &  \frac{i}{\sqrt{2}} (\dot\theta \mp \frac12\Omega_q) & \frac12\Omega
\end{array}\right] ,
\ee
has SU(2) symmetry and can be reduced to an effective two-state Hamiltonian \cite{Vitanov1997bec,Randall2018}.
However, this is not necessary here because for the pulse shapes of Eq.~\eqref{shapes-sech}, $\H_a^{L,R}$ contains the same time-dependent function $\sech^2(t/T)$ in all of its elements.
After the change of the independent variable $t \to \theta(t)$, $\H_a^{L,R}$ becomes constant and readily solved.
The propagator reads $\U_a = \exp[-i \int_{-\infty}^{\infty} \H_a^{L,R} (t) \dt]$.
Its explicit calculation is straightforward but the final expression is too cumbersome to be presented here.
The probability for transition $\ket{1} \to \ket{3}$ is equal to the probability to remain in the dark state $\ket{\chi_0} = \cos \theta \ket{1} - \sin\theta \ket{3}$ because $\ket{\chi_0}\to \ket{1}$ as $t\to -\infty$ and  $\ket{\chi_0}\to -\ket{3}$ as $t\to\infty$.
Therefore the population of the bare molecular state $\ket{3}$ reads $P_3 = P_0 = |(\U_a)_{22}|^2$, or explicitly,
\bse
\begin{align}
P_3^L &= 1, \\
P_3^R &= \left[ 1- \frac{ 2 \sin^2 \left( \frac12 \pi \sqrt{(A/4) ^2 + 1} \right)} {(A/4) ^2 + 1} \right]^2 . \label{P3-sech2}
\end{align}
\ese
The population $P_3^R$ vanishes when the P and S pulse area is $A \approx 1.017\pi$.
Then the contrast between L and R handedness is maximal.

\begin{figure}[tb]
\includegraphics[width=0.85\columnwidth]{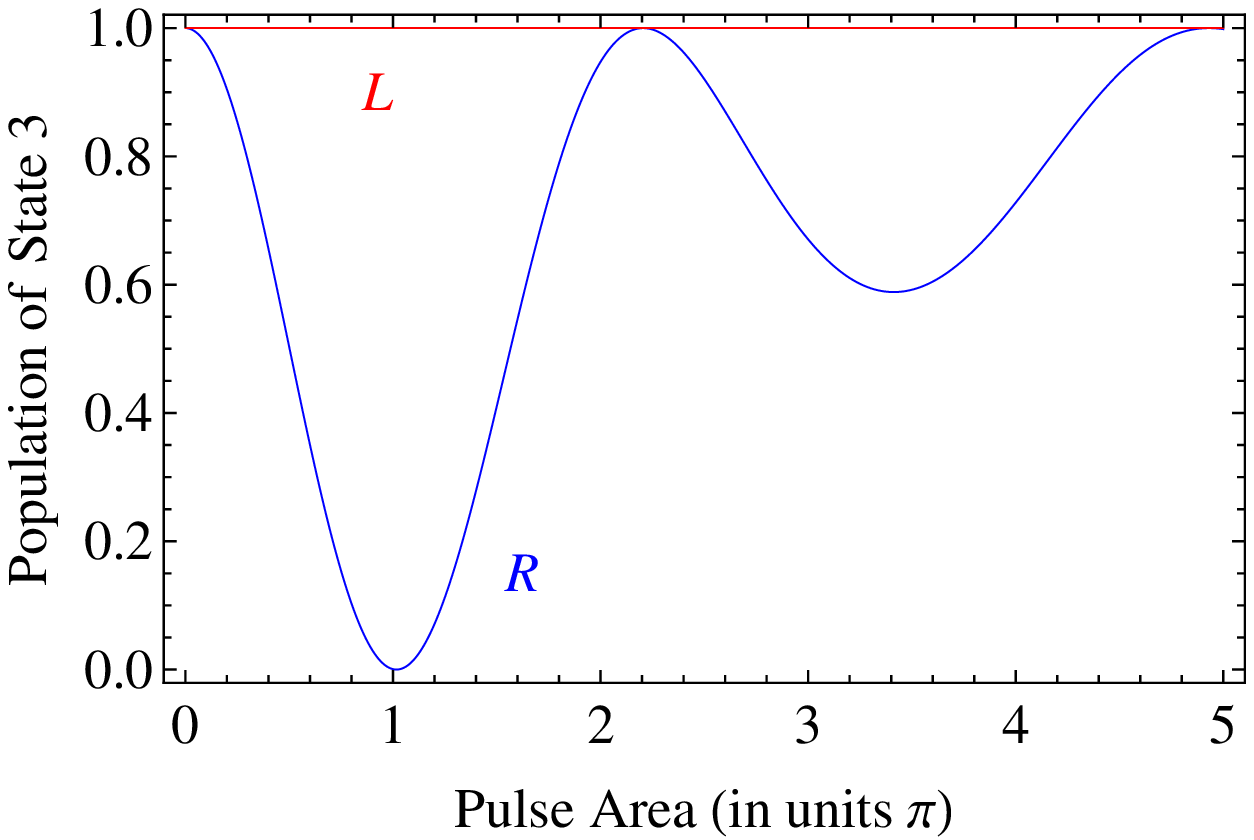}
\caption{
Population $P_3$, Eq.~\eqref{P3-sech2}, of state $\ket{3}$ vs the area $A$ of the P and S pulses.
The population $P_3$ for L handedness is exactly equal to 1 due to the ``shortcut to adiabaticity'', whereas it oscillates for R handedness due to the doubled nonadiabatic coupling.
As the pulse area $A$ increases, the adiabaticity improves and hence the R population also tends to 1.
The largest contrast between the signals occurs for $A\approx 1.017\pi$, when the R signal vanishes.
}
\label{fig:P3-sech}
\end{figure}

The population \eqref{P3-sech2} of state $\ket{3}$ for the analytic model \eqref{shapes-sech} is plotted vs the P and S pulse area $A$ in Fig.~\ref{fig:P3-sech}.
For the L handedness, $P_3=1$ due to ``shortcut to adiabaticity'' effect, while it oscillates for the R handedness.
The largest contrast between the L and R signals occurs for $A\approx 1.017\pi$, where the R signal vanishes.
At this value of the pulse areas the chiral resolution is ideal: a nonzero signal from state $\ket{3}$ unambiguously indicates a certain handedness while no signal indicates the opposite one.
